\documentclass[12pt]{iopart}
\usepackage{graphicx}
\begin{document}
\title{Charged surfaces and slabs in periodic boundary
  conditions}

\author{M J Rutter}

\address{TCM, Cavendish Laboratory, JJ Thomson Avenue, Cambridge,
  CB3~0HE, UK}

\begin{abstract}
  Plane wave density functional theory codes generally assume
  periodicity in all three dimensions. This causes difficulties when
  studying charged systems, for instance energies per unit cell become
  infinite, and, even after being renormalised by the introduction of
  a uniform neutralising background, are very slow to converge with
  cell size. The periodicity introduces spurious electric fields which
  decay slowly with cell size and which also slow the convergence of
  other properties relating to the ground state charge density. This
  paper presents a simple self-consistent technique for producing
  rapid convergence of both energies and charge distribution in the
  particular geometry of 2D periodicity, as used for studying
  surfaces.
\end{abstract}

\maketitle

\section{Introduction}

Plane-wave electronic structure codes generally enforce periodicity in
all three dimensions, and this periodicity applies not just to the
structure, but also to the calculated potential.  Whilst neutral bulk
crystals have this periodicity, systems such as surfaces and slabs are
of lower dimensionality. To study these systems in such codes, a
region of vacuum is used to separate the system from its fictitious
periodic images. As the extent of the vacuum is increased, in many
cases properties converge to the values that they would have in the
absence of the imposed periodicity. This convergence is often slow, so
corrections to accelerate the convergence have been derived for
various geometries\cite{MP95}. However, in the particular case of a
charged 2D system, the energy does not converge, but increases
linearly with vacuum extent.

Charged 2D systems are of interest in many areas, and have been the
subject of both experiment and of Density Functional Theory (DFT)
calculations. An inexhaustive list, with references to some recent
work, would include exfoliation \cite{Khazaei2018,Wu2019},
adsorption of gas molecules on surfaces\cite{Li2017,Bal2018},
and phase transitions in thin sheets\cite{Patil2019}. In all of these
areas there is a desire to study the effect of varying charge on the
process, and a need to resolve small energy differences accurately.

Various methods have been proposed for addressing the issues which
arise, including the use of a 2D Ewald summation in place of the usual
3D Ewald summation\cite{Yeh99}, truncated Coulomb
methods\cite{Jarvis97,Hine11},
extrapolation\cite{Komsa14,Nattino19}, minimum image Coulomb
potentials\cite{Hine11,MT99} and Green's function
approaches\cite{Otani06,Dabo08,Dabo11}. Some of these
techniques bring their own constraints, such as the Coulomb truncation
methods which require the extent of the vacuum region to be greater
than the extent of the non-vacuum region\cite{Rozzi06}. In this paper
a simple correction is described which can be applied to any code
which uses the standard 3D Ewald summation, and which greatly
accelerates the convergence of the energy other properties of the
wavefunction with vacuum size in charged systems.

When computing charged systems using periodic boundary conditions, the
periodicity causes an extra, unphysical electric field to arise, and
this perturbs the charge density from what it would have been in the
aperiodic case\cite{Krukowski13}. This is most noticeable if the
system is highly polarisable, for example, a metal. That the charge
density is significantly perturbed from the aperiodic ground state
means that simple \textit{post hoc} corrections will not be highly
accurate as they will be unable to correct the charge density,
although more sophisticated \textit{post hoc} schemes which estimate
the value of $\epsilon_r$ can improve
this\cite{Komsa13,Neugebauer18}. The problem is not confined to
plane-wave codes, for other codes also find it convenient to use
methods which force the potential to be periodic, and thus require
corrections when modelling systems in which it should be
aperiodic\cite{Hine11,Needs96}.

This paper discusses \textit{post hoc} corrections which merely
correct the energy, but not forces or density-dependent properties,
and it also introduces a self-consistent correction which not only
gives a superior correction to the energy and could be used to correct
forces, but the self-consistent correction also corrects the density
and thus improves the convergence of density-dependent properties.

\section{Theory}

A sheet carrying a uniform charge per unit area of $\sigma$ produces
an electric field on each side of magnitude
$0.5\sigma/\epsilon_0$, assuming symmetric boundary
conditions.

\begin{figure}
  \begin{center}
    \includegraphics[width=0.6\textwidth]{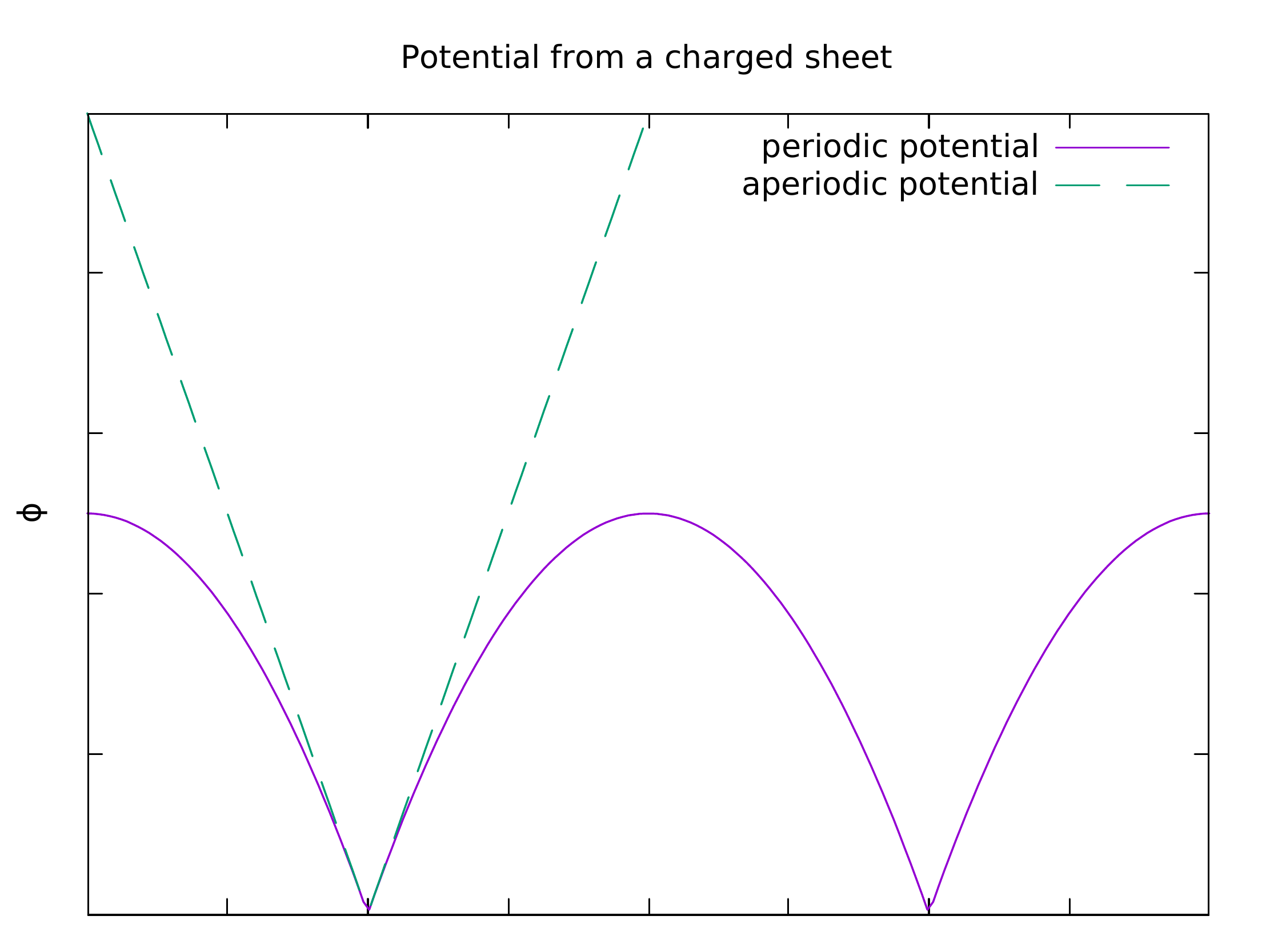}
  \end{center}
  \caption{The potential arising from a uniformly charged sheet with
    and without periodic boundary conditions. Two repeat units
    are show in the periodic case.}
  \label{fig:sheet}
\end{figure}

When such a sheet is modelled in a standard plane wave code, with
periodicity in all three dimensions imposed, the resulting potential
is as shown in figure~\ref{fig:sheet}. In the vacuum region where the
aperiodic system correctly has a constant field and a linear
potential, the periodic system has a linear field and a quadratic
potential. This can be understood as arising from the uniform
compensating charge placed throughout the cell, usually referred to as
`jellium'.

The method generally used in plane wave codes of dividing the
electrostatic interaction into the three parts of ion-ion,
electron-electron and ion-electron results in three infinite
terms. However, if the $g=0$ reciprocal space component of the charge
density is removed from each of these, not only do they become finite,
but the three removed terms cancel exactly for an uncharged system. So
the sum is always performed with the assumption that the total ionic
and electronic charges are equal. If the original total charges were
not equal, as would be the case in a charged system, then within the
calculation the total charges are treated as equal. This is done,
implicitly, by ignoring the $g=0$ components of the densities, which
is equivalent to adding a uniform compensating charge to the system to
make it neutral\cite{Leslie85}.

As can be seen from figure~\ref{fig:sheet}, in the immediate vicinity
of the sheet, the potential and field is the same in the periodic and
aperiodic cases. However, in the periodic case the field falls below
the aperiodic value in a linear fashion with increasing distance from
the sheet.

\begin{figure}
  \begin{center}
    \includegraphics[width=0.6\textwidth]{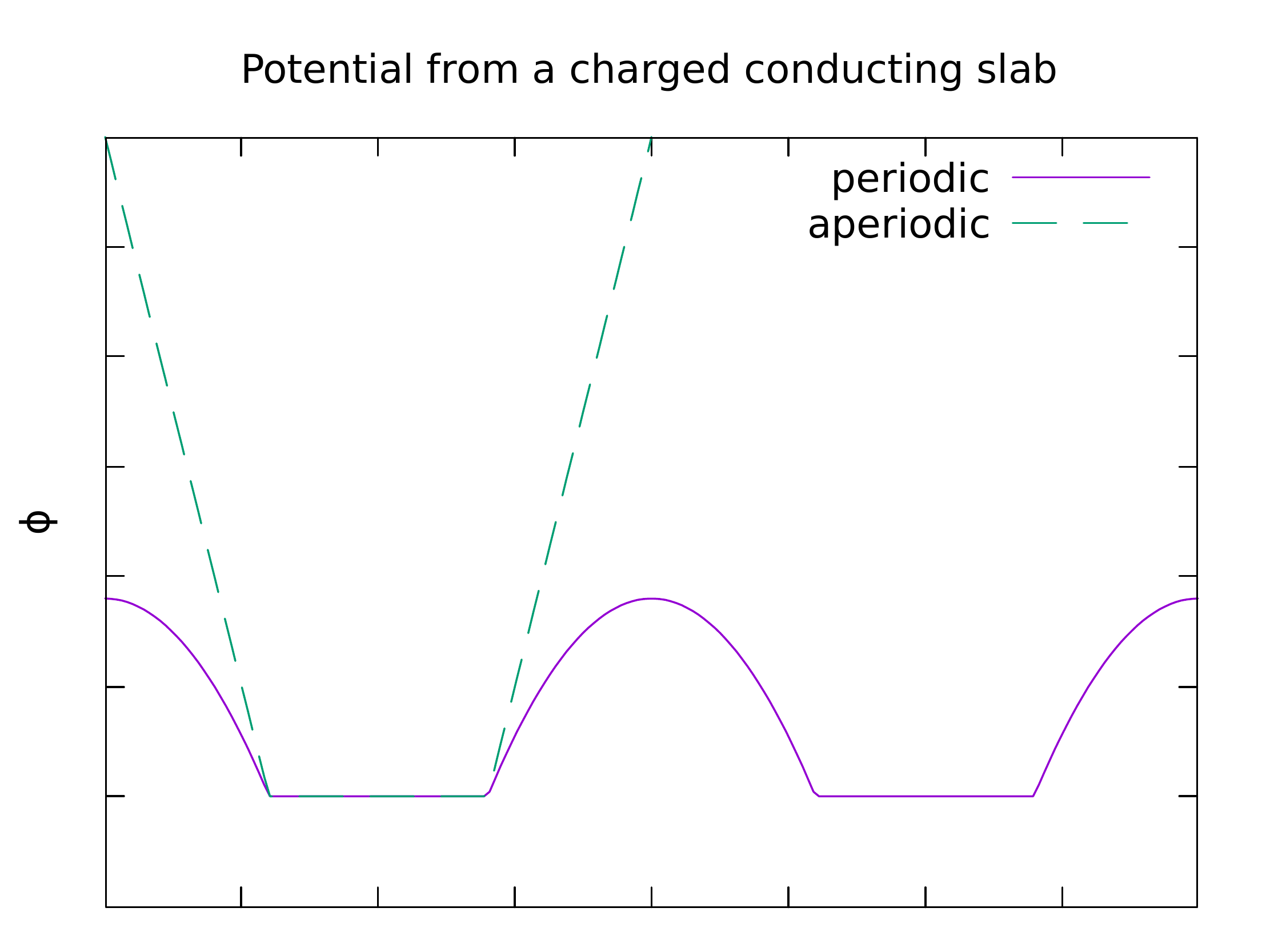}
  \end{center}
  \caption{The potential arising from a uniformly charged conducting
    slab with and without periodic boundary conditions. The thickness
    of the slab is 40\% of the periodic repeat distance in the
    direction normal to the slab.}
  \label{fig:slab}
\end{figure}

If, instead of a thin sheet, the charged body is a conducting slab of
significant thickness, the situation changes to that shown in
figure~\ref{fig:slab}. 

In this case, in the aperiodic case, the charge rests solely on each
surface of the slab, giving there a charge density of $0.5\sigma$, and
the field strength is as before. Within the conducting slab the
potential must be constant.

In the periodic case for the charged conducting slab, again the
potential within the conducting slab must be constant, and in the
vacuum region the quadratic term must be the same as for the thin
charged sheet, as the density of the jellium does not depend on the
thickness of the slab. If $z$ is the axis perpendicular to the slab,
then the electrostatic potential, $\phi$, is a function of $z$ only,
and, in the vacuum region $d^2\phi/dz^2=\rho_j/\epsilon_0$ where
$\rho_j$ is the charge density due to the neutralising jellium. In
this case the surface charge density on each side of the slab is no
longer $0.5\sigma$. It is reduced because some of the excess charge of
the slab must move to neutralise the jellium charge in the slab region
where the potential must be constant, and hence the total charge
density, including the jellium, must be zero. Thus the surface charge
density is reduced by a factor of the width of the vacuum region
divided by the periodic distance normal to the surface. The surface
field is reduced by the same factor.

\subsection{\textit{Post hoc} energy corrections}

If one considers a charged sheet, placed at the origin,
then, in the periodic case, the potential as a function of $z$ is

\begin{equation}
\phi=\frac{q}{2\epsilon_0A}|z|-\frac{q}{2\epsilon_0V}z^2-\frac{qc}{12\epsilon_0A} \label{eqn:phi_pbc}
\end{equation}

where $A$ is the cell's area in the plane of the slab, $c$ the repeat
distance perpendicular to the slab, $V=Ac$ its volume, and $q$ the
charge per unit cell of the sheet. The constant term arises because
codes set the average electrostatic potential (the reciprocal space
$g=0$ term) to zero. The energy of the sheet in this potential is just
$0.5q\phi(0)$, the factor of a half avoiding the double-counting
associated with the sheet being in the potential which it itself has
produced, which is

\begin{equation}
U=-\frac{q^2c}{24\epsilon_0A} \label{eqn:ph_Eq}
\end{equation}

This term is linear in $c$, the cell length normal to the slab, and so
is best removed by extrapolating $c$ to zero.

This expression can also be derived by integrating the energy density of
the electric field, $0.5\epsilon_0E^2$. In the half of the cell from
$z=0$ to $z=c/2$ the field, $E$, is given by differentiating
equation~\ref{eqn:phi_pbc},

\begin{equation}
  E(z)=\frac{q}{2\epsilon_0A}-\frac{qz}{\epsilon_0V} \label{eqn:ph_field}
\end{equation}

and then, considering the total contribution from the two equivalent
half cells,

\begin{equation}
U=2\int_0^{\frac{c}{2}}0.5\epsilon_0E^2\,A\, dz \label{eqn:field_int}
\end{equation}

which results in the same expression for the energy.

To find the next order term in the \textit{post hoc} correction to the
energy, consider splitting the thin charged sheet above into two
sheets, each of charge $q/2$, and separated by a distance $2d$. There
is an energy term which represents moving the charge up the linear
term of the `V'-shaped potential. However, as this linear part of the
potential is independent of $c$, so too is this energy, and, as a
constant offset in the energy, it can be ignored.

There is a contribution to the energy arising from the quadratic term in
equation~\ref{eqn:phi_pbc}. This is

\begin{equation}
U=-\frac{q^2}{4\epsilon_0V}d^2 \label{eqn:E1ph}
\end{equation}

with again a factor of two from double-counting\cite{BAC09}.
Furthermore there is an energy arising because the average of the
potential has now changed, so the constant term in
equation~\ref{eqn:phi_pbc} has to change to keep the average potential
zero. To first order, this requires an extra constant term of
$qd^2/(2\epsilon_0V)$. With the usual comments about overcounting,
this gives rise to a second contribution to the energy, identical in
form and sign to equation~\ref{eqn:E1ph}. If one further defines a
quadrupole moment $Q_{cc}=qd^2$, then the total \textit{post hoc}
energy correction is

\begin{equation}
U=-\frac{q^2c}{24\epsilon_0A}-\frac{qQ_{cc}}{2\epsilon_0V} \label{eqn:ph_EqQ}
\end{equation}

\subsection{Self consistent corrections}

In order to produce results which converge rapidly with vacuum size,
it is necessary to ensure that fields and charge distributions in and
near the slab are independent of the width of the vacuum region. This
can be achieved by adding a correcting potential which removes the
quadratic potential introduced by the jellium. Such a potential will
have a discontinuity in its derivative in the middle of the vacuum
region, and can be considered to represent a compensating charged
sheet, of charge density $-\sigma$, in the middle of the vacuum
region. It is thus similar to the method of Lozovoi and
Alavi\cite{Lozovoi03} who used a Gaussian-smeared sheet in the centre
of the vacuum region.  When considered as a compensating charge, the
net charge of the cell becomes zero, and no jellium is introduced into
the system. In order to achieve this, the potential which has to be added is

\begin{equation}
\phi_{\textrm{corr}}=\frac{qz^2}{2\epsilon_0V}-\frac{qc^2}{24\epsilon_0V} \label{eqn:phi_sc_corr}
\end{equation}

which will result in a total electrostatic potential of

\begin{equation}
\phi_{\textrm{tot}}=\frac{qc|z|}{2\epsilon_0V}-\frac{qc^2}{8\epsilon_0V} \label{eqn:phi_corr}
\end{equation}

which is identical to the electrostatic potential for the aperiodic
case in figure~\ref{fig:sheet}.

Introducing this correction to the potential results in an additional
energy correction of

\begin{equation}
U=-\frac{q^2c}{12\epsilon_0A} \label{eqn:E_sc}
\end{equation}

This arises from noting that this potential arising from the
fictitious charged sheet will be allowed to act on the electrons and
ions, but to include all the energy terms of the system including the
charged sheet, the extra potential must also be allowed to act on the
sheet itself, subject to a factor of a half to avoid double
counting. The charged sheet is placed at $z=c/2$, at which point the
value of $\phi$ as given by equation~\ref{eqn:phi_sc_corr} is

\begin{equation}
\phi_{\textrm{corr}}(\frac{c}{2})=\frac{qc^2}{12\epsilon_0V}
\end{equation}

This gives an extra energy term of

\begin{equation}
U=\frac{q^2c}{24\epsilon_0A} \label{eqn:U1}
\end{equation}

There remains one final correction to the energy. As the vacuum region
is expanded, the energy of the cell will increase due to the energy
density of the electric field in the vacuum. The field remains
constant with increasing vacuum size, but its volume increases. This
argument gave equation~\ref{eqn:field_int}, but, whereas in the case
of the \textit{post hoc} correction the field was given by
equation~\ref{eqn:ph_field}, in the self-consistent case the field is
now simply $\sigma/2\epsilon_0$ or $q/2\epsilon_0A$, as the field is
now identical to what it would be in the aperiodic system. So after
the self consistent correction the integral evaluates to
$q^2c/(8\epsilon_0A)$. This is not an energy term that is desired in
the final result, so it must be subtracted, and the term from
equation~\ref{eqn:U1} added. Thus the total correction is

\begin{equation}
U=-\frac{q^2c}{12\epsilon_0A}
\end{equation}

as stated in equation~\ref{eqn:E_sc}.

This energy can either be included explicitly, or by adding a constant
to equation~\ref{eqn:phi_sc_corr}. The latter approach has the effect
of changing the sign of the constant term in that equation.

\section{Non-uniform systems}

The above theory has been developed by considering uniformly-charged
sheets. A real surface does not have a uniform charge density or
potential, though it can be shown that the long-range effects of
non-uniformity in either the charge density or the potential decay
rapidly with increasing vacuum extent.

Consider an isolated thin sheet with a non-uniform charge distribution
such that the potential close to its surface is sinusoidal in the
plane of the sheet. In the vacuum region, Gauss's Law states that the
potential must obey $\nabla^2\phi=0$. Choosing $x$ to be the direction
of sinusoidal variation, the solution is

\begin{eqnarray}
  \phi & =  & a\exp(ikx-kz) + bz + c \\
  \nabla^2\phi & = & a (-k^2+k^2) \exp(ikx-kz) = 0
\end{eqnarray}

So any Fourier component of $\phi$ parallel to the slab decays
exponentially away from it (discounting the unphysical mathematical
possibility of an exponential increase). An arbitrary 2D-periodic
charge density on a sheet can be considered as the sum of Fourier
components, each of which must decay exponentially in the vacuum
region. The slowest-decaying component will be that with the smallest
$k$ value, i.e the longest wavelength. As $\phi$ must have the
periodicity of the unit cell, its periodic part will decay in the
vacuum region at least as fast as $\exp(-2\pi z/a)$ where $a$ is the
longer of the two cell axes in the plane of the slab. And as the potential
mediates the interaction between the system's periodic images, the difference
between the interactions from a uniformly charged slab, and a
periodically charged one, will decay similarly with increasing vacuum
extent. This confirms statements based on numerical observations about
the required vacuum extent for well-converged calculations\cite{Yeh99}.

\section{Non-symmetric systems}

The systems considered so far all have sufficient symmetry to cause
their dipole moment to be zero about the centre of the slab. However,
the theoretical approach presented in this paper is readily
generalisable to systems lacking this symmetry. In a charged system it
is possible to pick an origin about which the dipole moment is zero,
and then this self-consistent correction can be applied centred on
this origin\cite{MP95}.

A shift to the point about which the dipole moment is zero requires a
shift of $p/q$, where $p$ is the dipole moment and $q$ the net
charge. Applying this shift to equation~\ref{eqn:phi_sc_corr} leads to
the emergence of a linear term in the potential,
$pz/\epsilon_0V$. This is the expected correction for the
dipole-dipole interaction between uncharged slabs as given by
Neugebauer and Scheffler\cite{Scheffler92}, showing an equivalence
between shifting co-ordinates to eliminate the dipole moment, and
explicitly including it.

This generalisation enables a correction to the forces on the ions to
be obtained by differentiating the energy correction with respect to
the ionic positions, although this extension is not further discussed
here.

\section{Validation calculations}

To verify the theory presented above, calculations were performed on a
system comprising of a varying number of graphene sheets stacked in
the ABAB sequence. These systems are metallic. A calculation was also
performed on a non-metallic system which lacks inversion symmetry, a
slab of SiC. No atomic relaxation was performed in these calculations,
as only the electronic minimisation was of interest. In the graphene
calculations the in-plane lattice parameter was set to 2.46\AA{}, and
for SiC to 3.08\AA{}.

Firstly results are presented for a single layer of graphene, with a
charge of +2e per unit cell (pair of atoms). This system acts as a
thin, charged sheet.  Secondly results are presented for a four layer
system of graphene, again with a charge of +2e per unit cell. This
system acts as a charged conducting slab, with the graphene occupying
the majority of the cell when the vacuum separation is small. Finally
results are presented for a thin slab of SiC, cut along its
$\lbrace$0001$\rbrace$ planes, as an example of a system which lacks
inversion symmetry.

The calculations were all performed with a high cutoff of 550eV, as
the positive charge tends to make orbitals more compact requiring a
larger basis set, and using the Francis-Payne correction for a finite
basis set\cite{FP90}. The validity of the corrections presented will
not depend on the exchange-correlation functional used, so these
calculations used the computationally-inexpensive LDA.

The \textsc{Castep}\cite{CASTEP} code was used for all the
calculations. The \textit{post hoc} corrections were performed using a
modified version of c2x\cite{c2x}, whereas the self consistent
corrections used a version of \textsc{Castep} modified for this purpose.

As well as considering the convergence of the total energy, the
convergence of one component of the quadrupole moment is also
considered, the same component which appears in the \textit{post hoc}
energy correction, equation~\ref{eqn:ph_EqQ}. The convergence of this
moment is another measure of how well converged the electron density
is, and is a property which is not readily corrected by \textit{post
  hoc} methods. This is calculated in the direction perpendicular
to the slab, after averaging over the two orthogonal directions, as

\begin{equation}
Q_{cc}=\int_0^c (z-0.5c)^2\rho(z)\,dz
\end{equation}

Where $c$ is the extent of the cell in the $z$ direction perpendicular
to the slab, and $\rho(z)$ the total charge density after averaging
over the directions parallel to the slab. The slab is centred at
$z=c/2$ with $z=0$ (or $z=c$) being the centre of the vacuum region.

\subsection{Single Layer of Graphene}

\begin{figure}
  \begin{center}
    \includegraphics[width=0.4\textwidth]{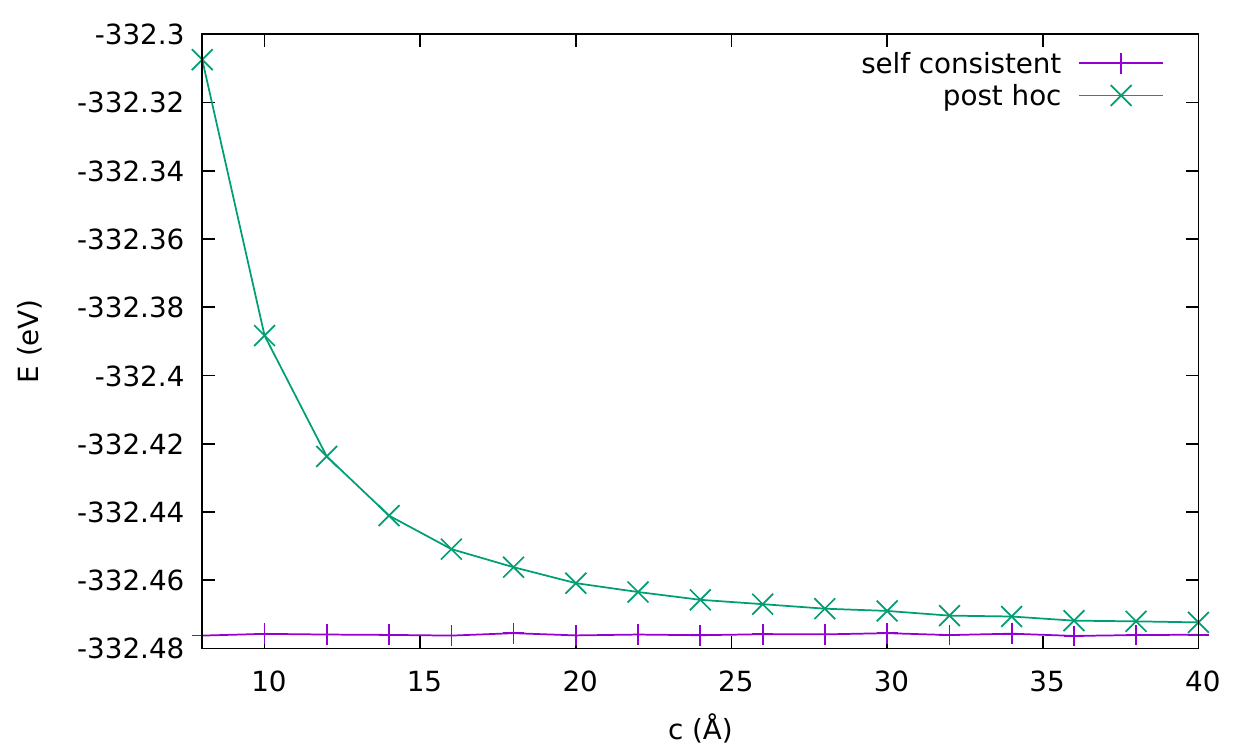}%
    \hspace{0.1\textwidth}%
    \includegraphics[width=0.4\textwidth]{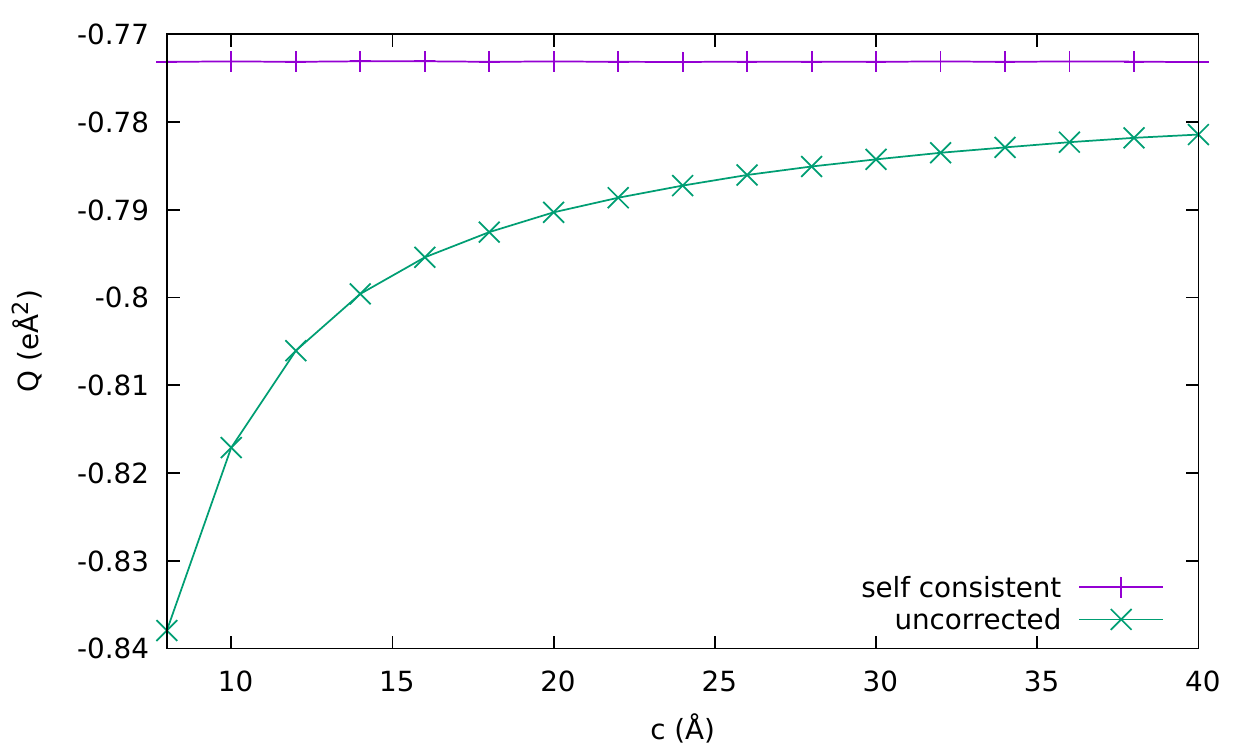}
  \end{center}
  \caption{The energy (left) and quadrupole moment (right) from a
    single layer of graphene with a charge of +2e per unit cell (2
    atoms) as the cell length (vacuum separation) is varied. For the
    energy a \textit{post hoc} correction is shown as well as a
    self-consistent correction. For the quadrupole, no \textit{post
      hoc} correction was applied.}
  \label{fig:1layer_EQ}
\end{figure}

A single layer of positively-charged graphene is well modelled by a
thin sheet of charge. With a charge of +1e per atom, the $\pi$
orbitals will be unoccupied, and the charge density is close to two
dimensional. The \textit{post hoc} corrections, which would be exact
if the change in the field caused by the periodicity did not cause a
change in the charge distribution, should therefore be quite good.

Figure~\ref{fig:1layer_EQ} shows the convergence of the energy, after
the \textit{post hoc} correction, and also the convergence of the
quadrupole moment. Both are compared to calculations with a
self-consistent correction.

No \textit{post hoc} correction can be applied to the quadrupole
moment without extra information such as the thickness and
polarisability of the system. The geometry of the system is such that
the variation of the quadrupole moment is not great, but the
application of the self-consistent correction gives a very constant
value. The variation of the quadrupole moment with the self-consistent
correction shows no obvious trend, just random noise, as it varies
between -0.77309 and -0.77318~e\AA$^2$. With a cell length of just
8\AA{} both the quadrupole moment, and the energy, are converged in
the self-consistent case.

\subsection{Four Layers of Graphene}

\begin{figure}
  \begin{center}
    \includegraphics[width=0.6\textwidth]{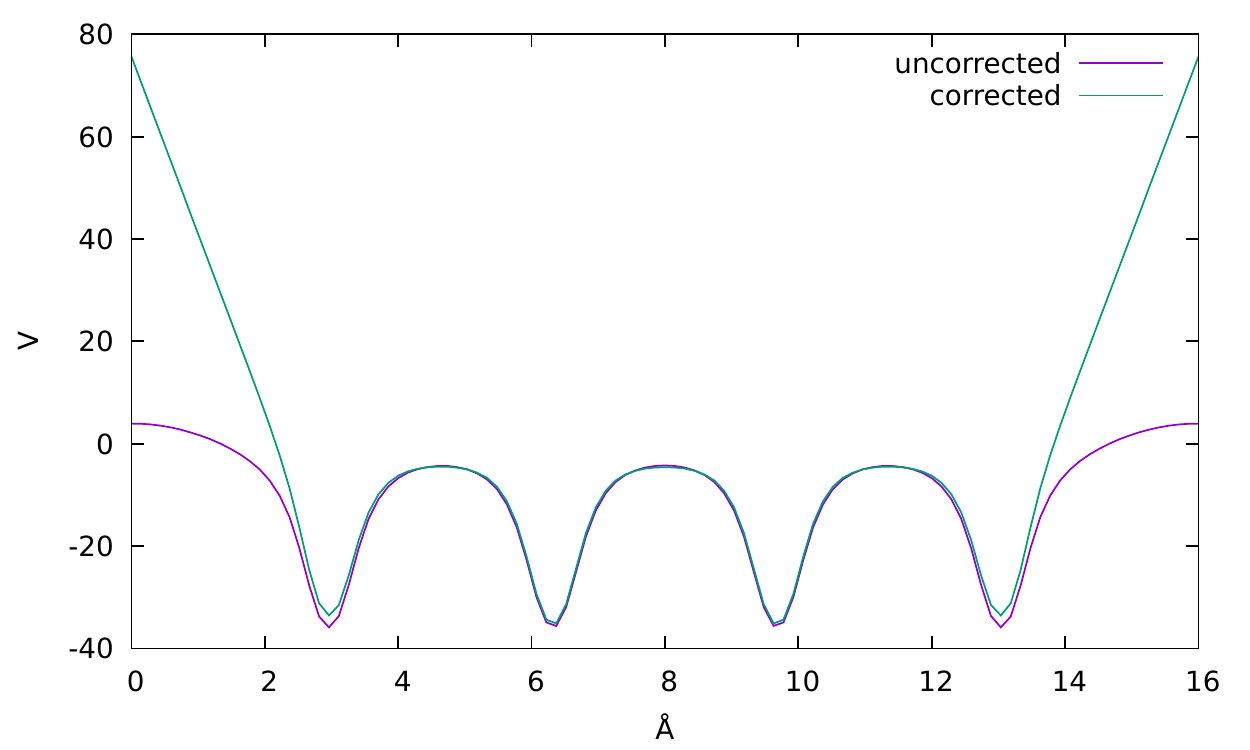}
  \end{center}
  \caption{The potential arising from four layers of graphene, with a
    charge of +2e per unit cell (8 atoms). A standard uncorrected DFT
    calculation is shown, along with the potential after the correction
    discussed in this paper. The uncorrected potential has been
    shifted to align the two potentials in the centre of the slab.}
  \label{fig:4layer_16A_pot}
\end{figure}

Four layers of graphene were stacked in an ABAB sequence with a
3.36\AA{} layer separation. The total extent of the slab is therefore
approximately 13.5\AA{}. Calculations were performed with the repeat
distance perpendicular to the slab varying from 16\AA{} to 40\AA{} in
2\AA{} steps. Figure~\ref{fig:4layer_16A_pot} shows the potential averaged over planes parallel to the slab when
the repeat distance is 16\AA{} in both the uncorrected case, and after applying
the self-consistent correction of equation~\ref{eqn:phi_sc_corr}.

If one ignores the oscillations in the potential within the slab
itself, this figure looks similar to figure~\ref{fig:slab}. The field
at the slab's surface is considerably suppressed in the uncorrected
calculation, and this will alter the surface properties.

\begin{figure}
  \begin{center}
    \includegraphics[width=0.4\textwidth]{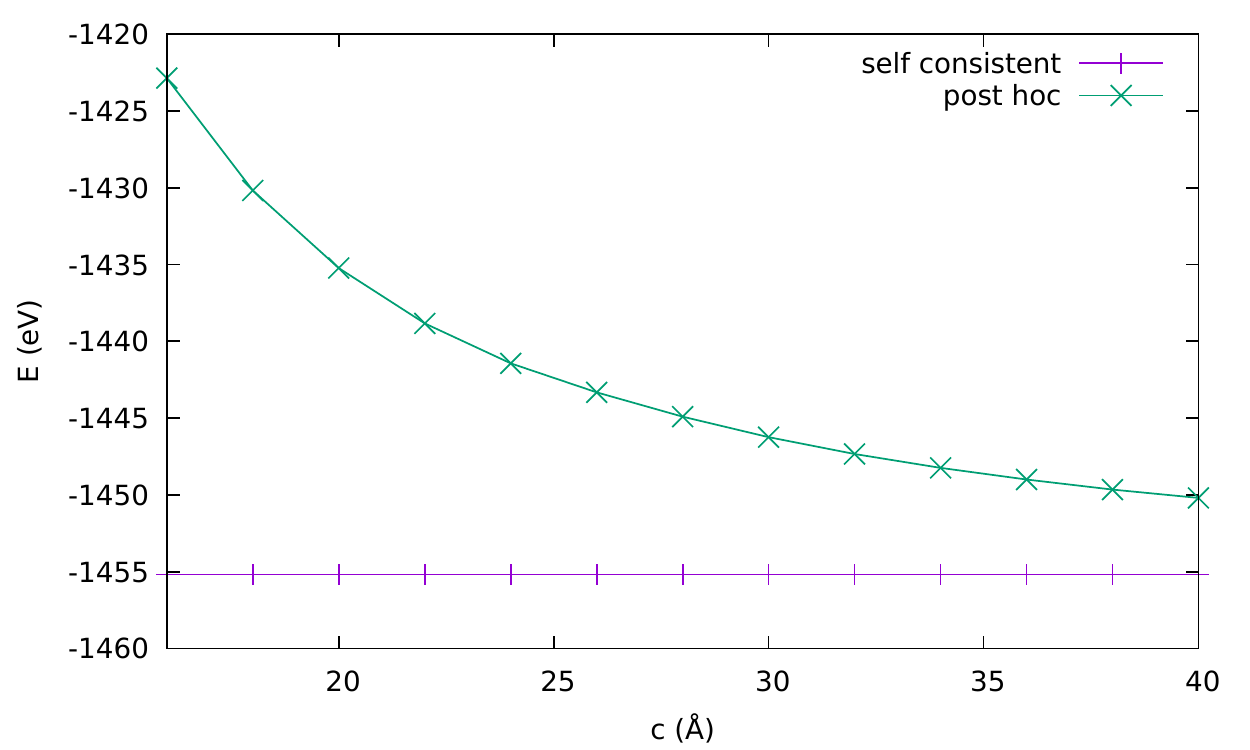}%
    \hspace{0.1\textwidth}%
    \includegraphics[width=0.4\textwidth]{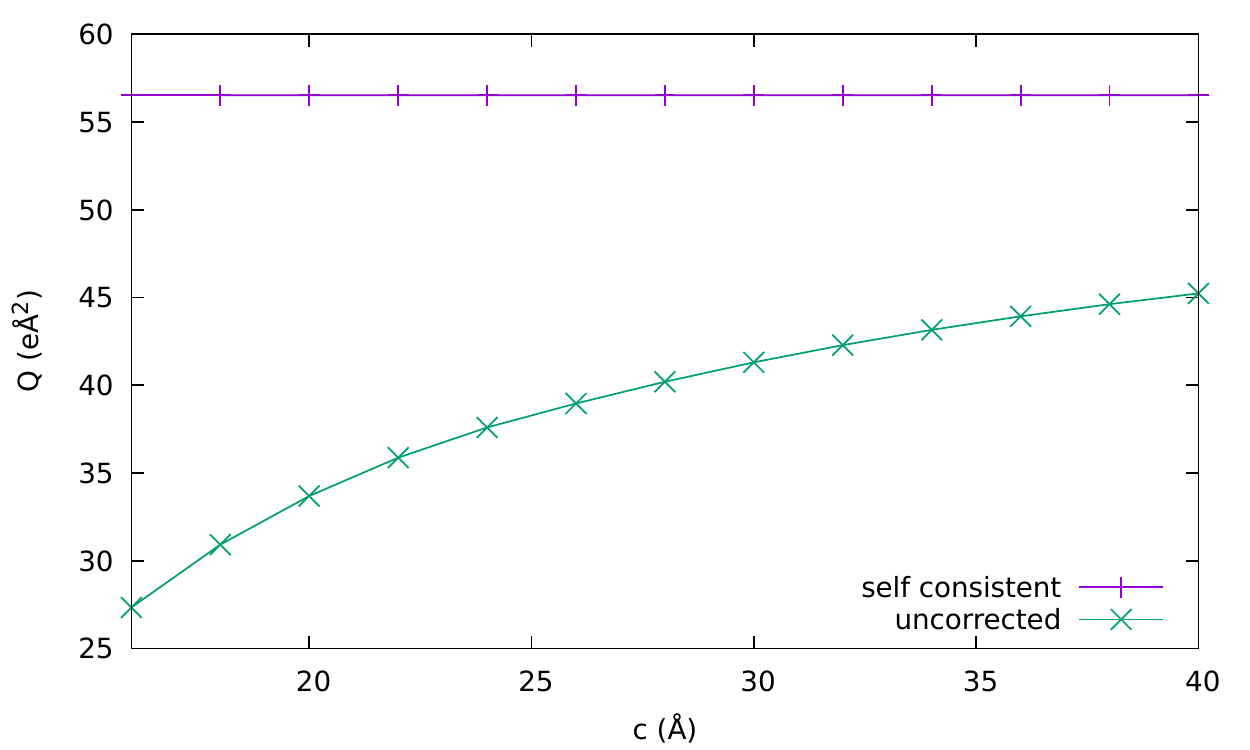}
  \end{center}
  \caption{The energy (left) and quadrupole moment (right) computed for
    four layers of graphene with a charge of +2e per unit cell (8
    atoms) as the cell length (vacuum separation) is varied. For the
    energy a \textit{post hoc} correction is shown as well as a
    self-consistent correction. For the quadrupole, no \textit{post
      hoc} correction was applied.}
  \label{fig:4layer_EQ}
\end{figure}

Figure~\ref{fig:4layer_EQ} shows the convergence of the energy of the
four layer graphene system, after the \textit{post hoc} correction,
and also the convergence of the quadrupole moment, and is thus the
four-layer equivalent of figure~\ref{fig:1layer_EQ}. The
self-consistent correction still shows very good convergence with cell
size. The \textit{post hoc} energy correction and uncorrected
quadrupole moment now show quite poor convergence. However, this is to
be expected for the following reason.

If one models the graphene slab as simply a conductor of thickness
$2L$ and charge $2q$, then, in the absence of periodicity, the charges
will lie on the two surfaces, and the quadrupole moment will be
$Q_{cc}=2qL^2$. Once periodicity is introduced, and with it its
neutralising jellium, the surface charges are reduced by a factor of
$1-\frac{2L}{c}$. It is hard to define precisely the thickness of a
slab of just a few atomic layers, but the thickness measured between
the surface nuclei is about 10\AA. So when $c$ is 40\AA, one would
expect the quadrupole moment to have reached about three quarters of
the value it would have when $c$ is infinity. The self-consistent
calculation produces values for $Q_{cc}$ which vary randomly between
56.522 and 56.538~e\AA$^2$, whereas the non self consistent
calculation gives $Q_{cc}$ as 45.226~e\AA$^2$ at $c=$40\AA. This is
almost exactly 80\% of the self-consistent value, but the simple
theory of this paragraph which predicted 75\% neglected the
contribution to the quadrupole moment of the bulk material, and the
setting of $2L=10$\AA{} is an underestimate. Thus it can be seen that
the quadrupole moment converges slowly with increasing vacuum
separation in the absence of any self-consistent correction, but
converges rapidly with the correction presented in this paper. For
this system, it is well converged when the vacuum separation is still
smaller than the slab thickness.

\subsection{A Non-Symmetric System}

\begin{figure}
  \begin{center}
    \includegraphics[width=0.6\textwidth]{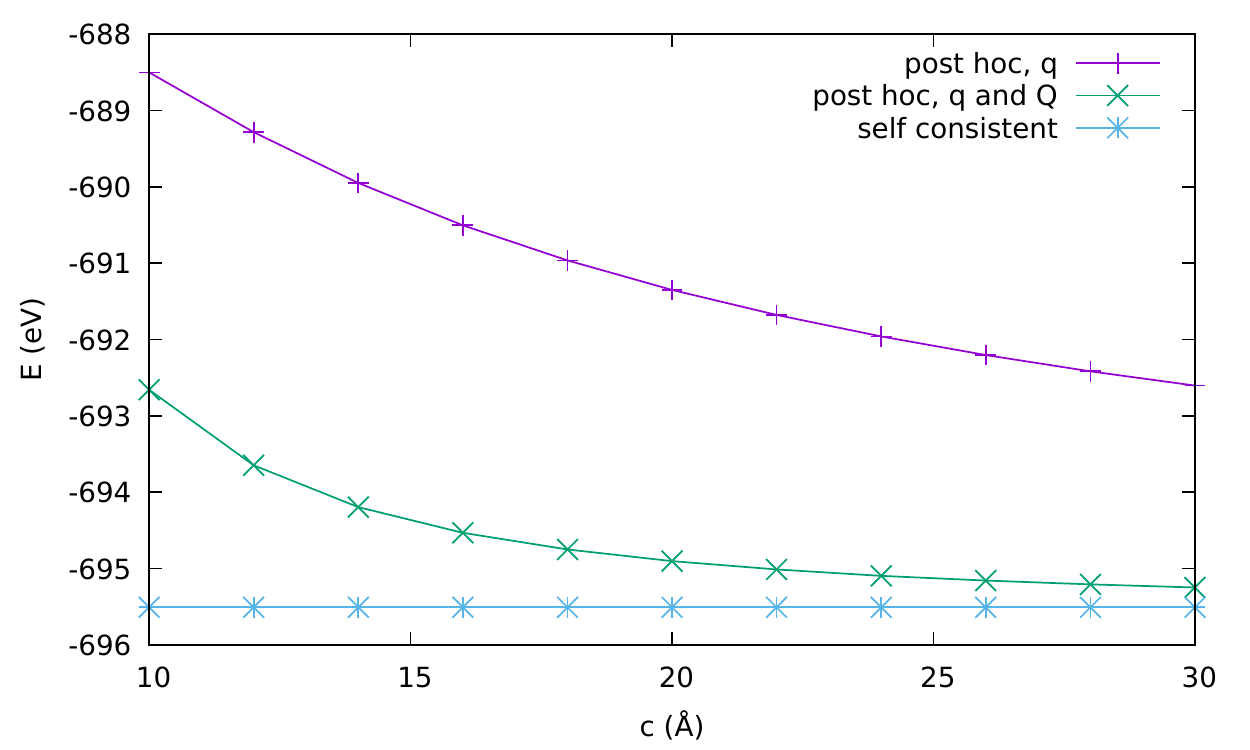}%
  \end{center}
  \caption{The energy from two layers of SiC with a charge of +2e per
    unit cell (4 atoms) as the cell length (vacuum separation) is
    varied. A simple \textit{post hoc} correction depending just on
    the charge is shown, a \textit{post hoc} correction including the
    quadrupole moment, and a self-consistent correction. The
    self-consistent line shows noise, and the variation between
  its greatest and least value is less than 0.5meV.}
  \label{fig:E_SiC}
\end{figure}

A slab of silicon carbide cut along its $\lbrace$0001$\rbrace$ planes
lacks inversion symmetry and, if uncharged, will have a dipole moment
perpendicular to the slab. For good convergence with cell size in the
neutral system a self-consistent correction for this dipole moment is
required\cite{RutterSiC}. In the following, two layers of a charged
system are considered, again with an overall charge of +2e per unit
cell. The distance between the planes of the centres of the nuclei at
the two surfaces (carbon on one surface, silicon on the other) was
just over 3.1\AA{}.

Calculations were performed in the same manner as for the graphene
sheets, save that the centre of the quadratic correcting potential was
no longer fixed at the centre of the slab, but allowed to move
self-consistently to the point about which the dipole moment of the
charged system was zero.

Figure~\ref{fig:E_SiC} shows the energy after removing the linear term
using the \textit{post hoc} correction of equation~\ref{eqn:ph_Eq},
the energy after using the better \textit{post hoc} correction of
equation~\ref{eqn:ph_EqQ} which includes consideration of the
quadrupole moment $Q_{cc}$, and also the energy from the
self-consistent calculation. The self-consistent energy shows no
systematic variation with $c$ in this range, and the extent of the
random variation is less than 0.5meV. Even when $c=30$\AA{}, the
better of the \textit{post hoc} corrections is still about 0.25eV from
the self-consistent result, showing that the self-consistent
correction has again achieved convergence with respect to the width of
the vacuum region at a much smaller cell size than the \textit{post
  hoc} corrections.

\section{Details of Implementation}

Two details of the implementation of the self-consistent correction
are worth comment. Firstly the systems chosen have a postive
charge. Care must be taken when working with negatively-charged
systems as the field in the vacuum region makes it increasingly
attractive for electrons to leave the bulk and to form a pool in the
middle of the vacuum region. Similar arguments apply to insulating
systems should the variation of the potential from the centre to the
surface of the slab exceed the bandgap.

\begin{figure}
  \begin{center}
    \includegraphics[width=0.65\textwidth]{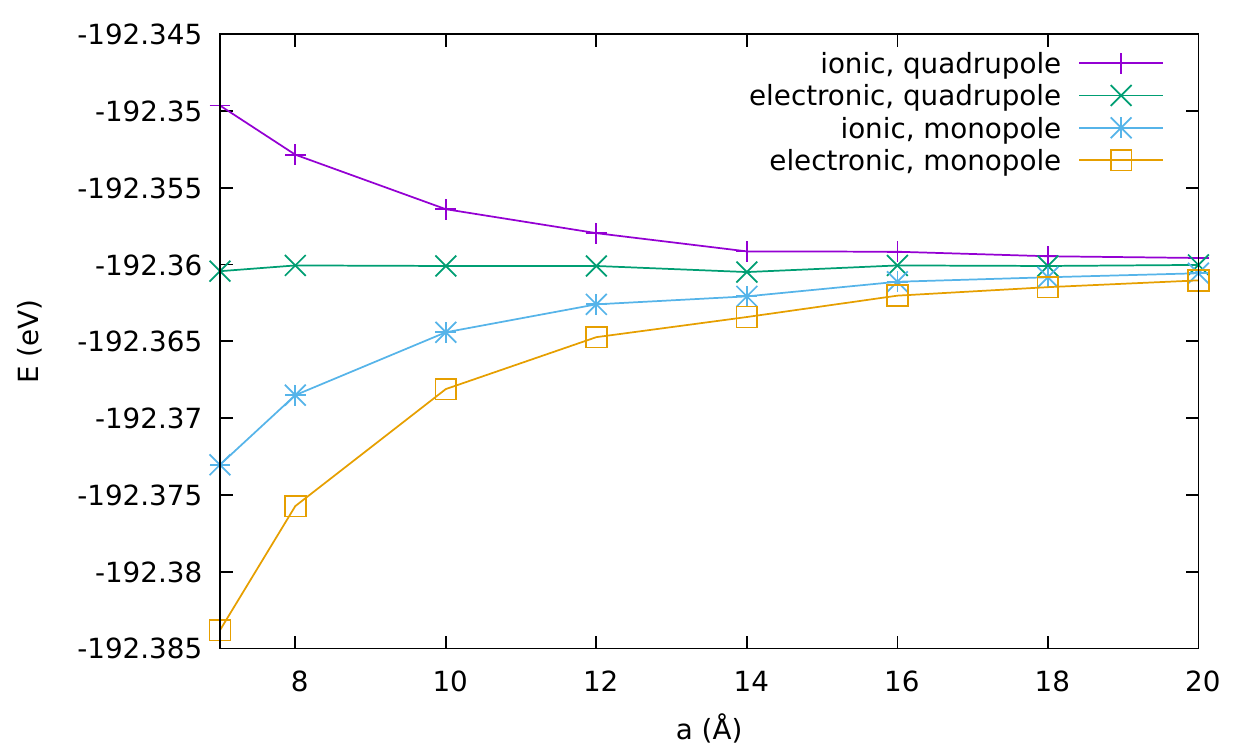}
  \end{center}
  \caption{The energy of an Li$^+$ ion in a cubic box as a function of
    cube size. Two different schemes for calculating the non-Coulombic
  energy are used, using the total ionic, and the total electronic,
  charge. Two different corrections for the charged system are used:
  the monopole term of the Makov--Payne correction, and both the
  monopole and quadrupole terms of that correction. The best
  convergence with size is seen with the full Makov--Payne correction,
  and the use of the ionic charge in the non-Coulombic energy.}
  \label{fig:Li}
\end{figure}

Secondly pseudopotential codes will have a term in their total energy
which arises from the non-Coulombic part of the $g=0$ components of
the pseudopotentials used. This gives rise to no fields, but to an
energy of

\begin{equation}
U=\frac{q}{V}\sum_{\textrm{ions}}V_{nc} \label{eqn:nc}
\end{equation}

Where $q$ is the total charge, $V$ the cell volume, and $V_{nc}$ the
non-Coulomb $g=0$ term of each ion's pseudopotential. This expression
is unambiguous in a neutral system, but in a charged system it matters
whether $q$ is the total electronic charge or the total ionic charge
(which can also be considered as the total charge from the electrons
and the jellium). For the corrections in this paper, as for those in
the original Makov and Payne paper\cite{MP95}, the total electronic
charge must be used. Historically this is the value used by
\textsc{Castep}\cite{PTAAJ92}, but recent versions of \textsc{Castep}
use the total ionic charge, as does Abinit\cite{Abinit}. The use of
the wrong value of $q$ here introduces an error in the energy which
decays as $1/c$ in this 2D geometry.

This point is illustrated by a calculation on a 0D system of a single
Li$^+$ ion in a cubic box of varying sizes. A 900eV basis set cut-off
was used. Figure~\ref{fig:Li} shows the energies after applying the
Makov--Payne correction using both the ionic and electronic
total charge in equation~\ref{eqn:nc}. It can be seen that, if one
merely includes the monopole term from the Makov--Payne correction,
then using the ionic charge in equation~\ref{eqn:nc} gives the better
result. However, on using both the monopole and quadrupole
corrections, the use of the electronic charge in equation~\ref{eqn:nc}
is seen to be much superior.

One can further strengthen this argument by considering a system of
Li$^{3+}$. Then only term in the energy will be the Ewald energy of
the ions, as all terms involving electrons must be zero. The variation
of the Ewald energy with cell size is exactly corrected by the
Makov--Payne formula, so there should be no other size-dependent term
in the energy. In other words, when there are no electrons,
equation~\ref{eqn:nc} should be independent of cell size, and so the
value of $q$ must be the electronic charge.


\section{Conclusions}

A self-consistent correction for modelling charged surfaces has been
presented. Unlike the \textit{post hoc} corrections, it produces rapid
convergence with vacuum size of the energy and of other properties
dependent on the wavefunctions or ground state charge density, such as
the quadrupole moment. The energy convergence is exponential in the
width of the vacuum region, and depends on the longest periodic
distance parallel to the surface. In contrast the \textit{post hoc}
corrections do not produce exponential convergence, and their
convergence rate depends on the thickness of the slab modelled. Though
a vacuum region must be present, in order to contain the discontinuity
in the potential, it is not subject to the constraint of being wider
than the slab's thickness, as methods such as truncated Coulomb
potentials require.

The corrections presented are easy to add to existing density
functional theory codes.

\section{Acknowledgements}

The author thanks Dr~John Biggins and Prof.~Volker Heine for helpful
discussions. This work was supported by EPSRC grant number
EP/P034616/1.

\section*{References}

\bibliographystyle{iopart-num}
\bibliography{charged_slabs}

\end{document}